\documentclass[a4paper]{article}
\usepackage{INTERSPEECH2019}

\usepackage{graphicx}
\usepackage{amssymb,amsmath,bm}
\usepackage{textcomp}

\usepackage{amsmath,graphicx}
\usepackage[font=small]{caption}
\usepackage{tabularx}
\usepackage{threeparttable}
\usepackage{subcaption}
\usepackage{dblfloatfix}
\usepackage{fixltx2e}
\usepackage{multirow}
\usepackage{booktabs}
\usepackage{balance}
\usepackage{amssymb}
\usepackage[compress]{cite}
\graphicspath{{./Figures/}}
\DeclareGraphicsExtensions{.pdf,.png,.eps}
\usepackage{epstopdf}

\ninept

\title{{EML} Online Speech Activity Detection for the Fearless Steps Challenge Phase-III}
\name{Omid Ghahabi, Volker Fischer}
\address{EML Speech Technology GmbH, Berliner Stra{\ss}e 45, 69120 Heidelberg, Germany}
 \email{(omid.ghahabi | volker.fischer)@eml.org}

\begin{document}

  \maketitle
  \begin{abstract}
  Speech Activity Detection (SAD), locating speech segments within an audio recording, 
  is a main part of most speech technology applications. Robust SAD is usually more difficult
  in noisy conditions with varying signal-to-noise ratios (SNR). 
  The Fearless Steps challenge has recently provided 
  such data from the NASA Apollo-11 mission for different speech processing tasks including SAD.
  Most audio recordings are degraded by different kinds and levels of noise varying within and between channels. 
  This paper describes the EML online algorithm for the most recent phase of this challenge. 
  The proposed algorithm can be trained both in a supervised and unsupervised manner and assigns speech and non-speech labels
  at runtime approximately every 0.1 sec. 
  The experimental results show a competitive accuracy on both 
  development and evaluation datasets with
  a real-time factor of about 0.002 using a single CPU machine.
  \end{abstract}
 \vspace{0.2cm} 

\section{Introduction}
\label{sec:intro}
Speech Activity Detection (SAD) is a fundamental signal processing step in almost
every speech processing application. Speech segments should be detected accurately in different environments with different 
levels and types of noise or in presence of other complicated non-speech acoustic events. The conditions could get even worse
when the audio signals are further degraded by transmission channel distortions.
All these conditions are summarized in the data provided by the Fearless Steps Challenge series\cite{joglekar2020fearless,
Hansen2019,Hansen2018}. 

The challenge aims to evaluate the performance of state-of-the-art speech
and language technologies on naturalistic audio recordings. The provided datasets are from 
the NASA Apollo-11 mission control recordings degraded by different types of noise, e.g., due to channel, system,  
attenuated signal bandwidth, transmission, cosmos, analog tape, and tape aging\cite{Hansen2018}. 
Additionally, the level of the noise varies both within and between different channels in a 25 dB range\cite{Hansen2018}.  
These conditions degrade significantly the speech detection accuracy and the robustness of  
traditional SAD algorithms. Phase III of this challenge is even harder than the previous phases 
in that the evaluation set is much more challenging and the non-scoring collar on the speech segment borders is
decreased from 0.5 sec to 0.25 sec.

Various SAD algorithms have previously been proposed for phases I and II in this challenge,
e.g.,\cite{vafeiadis2019two,kaushik2018speech
,gorin2020houston,lin2020dku,heitkaemperstatistical,gimenoconvolutional}. 
Most of the proposed algorithms and models are based on advanced deep learning techniques,
mainly Convolutional and Recurrent Neural Networks,
learning the statistical properties of speech and non-speech frames within the available training data.
On one hand, these techniques are usually too complicated and time-consuming for a SAD task. Note that SAD is just
a pre-processing stage for other speech processing applications like Automatic Speech Recognition, Speaker Diarization, etc.
So, it should be as little time-consuming as possible while being still accurate and robust enough.
On the other hand,
fully-supervised SAD may not work well on unseen environments which have not been in the training data.
Even though there are some domain adaptation techniques which adapt the decision threshold or the statistical models
to the new environment, they usually need the whole testing utterance for adaptation which is not feasible 
in real-time applications.

In this paper, we present the most recent EML online speech activity detection algorithm for the Fearless Steps challenge phase III. 
In the proposed algorithm, the speech and non-speech segments are modeled with two single vectors, one based on zero-order 
Baum-Welch statistics and the another based on a Deep Neural Network (DNN) transformation of first-order statistics
which are referred to as SAD embeddings in this paper.
In the testing phase, these two vectors are extracted every approximately 0.1 sec and compared to the model vectors at runtime.
Three main contributions are added to the previous version of the algorithm\cite{ghahabi_ESSV_VAD_2018}, to increase the accuracy 
and robustness for this challenging data. 
First, the feature vectors are transformed in two steps given the temporal contexts, to have
more discriminative and robust features. 
Second, both modeling SAD vectors and the decision threshold are temporally adapted to the new acoustic conditions at runtime.
Third, the SAD embeddings are used to capture complementary information for more reliable decisions.  

Experimental results show a competitive accuracy on both development and evaluation sets with much lower 
computational cost compared to other typical algorithms. The Decision Cost Function (DCF) values on the dev and eval sets
are, respectively, 2.20\% and 3.80\%, compared to the challenge baseline results with 12.50\% and 15.61\%. 

The rest of the paper is organized as follows. Section \ref{sec:challenge} briefly describes the 
datasets, the SAD task, and the baseline system in the challenge.
The proposed algorithm is described in more details in Section~\ref{sec:proposed-algorithm}.
 Section \ref{sec:results} analyzes the performance of the proposed SAD in terms of accuracy and computational cost.
Section \ref{sec:conclusions} summarizes the paper and discusses the future work.

\section{Fearless Steps Challenge}
\label{sec:challenge}
The Fearless Steps challenge series have been organized by UTDallas-CRSS since two years ago\cite{joglekar2020fearless,
Hansen2019,Hansen2018}. 
The latest phase (phase III) has also been co-organized by the National Institute of Standards and Technology (NIST).
The data is provided by NASA from Apollo-11 mission control recordings which are digitized from old tapes. 
The multi-channel and multi-speaker audio recordings are degraded by many different kinds of noise due to transmission channel, recording systems, 
attenuated signal bandwidth, cosmic noise, analog tape, and tape aging\cite{Hansen2018}.
Moreover, the level of the noise varies within and between recordings. There are different speech processing
tasks in the challenge but this paper focuses only on the SAD task. 
\subsection{Datasets}
\label{subsec:datasets}
Officially, 125 and 30 manually-labeled audio recordings, approximately 30 min each, have been provided for
training and development, respectively. Additionally, about 19,000 hours unlabeled data from the Apollo-11 mission could
be requested from the organizers and the participants have been allowed to use any other kind of data for training of their 
algorithms. The final evaluation has been performed on 68 recordings scored through the NIST platform. 
However, in this work we have used only the labeled data provided in the challenge for training of all parts of our algorithm.

\subsection{Task Description}
\label{subsec:SAD-task}
The SAD task aims to automatically locate speech segments in an audio recording. 
For evaluation, the start and end times of speech and non-speech 
segments from a hypothesized system output are compared to human annotated start and end times. 
This will result in two probabilities of errors namely False Positive ($P_{FP}$), detecting speech where there is no 
speech, and False Negative ($P_{FN}$),
not detecting speech where there is speech. 
The Detection Cost Function (DCF) is defined as a weighted sum of these two probabilities as follows,  
\begin{equation}
DCF(\theta) = 0.75\times P_{FN}(\theta) + 0.25\times P_{FP}(\theta)
\label{Eq:DCF}
\end{equation}
where $\theta$ is the system decision threshold. As it can be noticed in Eq.~\ref{Eq:DCF}, loosing speech segments is 3 times 
more costly than incorrectly detection of non-speech segments as speech. DCF given the hypothesized system threshold is used as 
the main evaluation metric for this task. A 0.25 sec collar at the beginning and end of ground truth speech segments
will not be scored, taking into account inconsistencies in human annotations. This collar was 0.5~sec in the previous 
phases of the challenge.  

\subsection{Baseline}
\label{subsec:SAD-baseline}
The baseline system used in the challenge is based on a Combo-SAD system developed primarily for the spontaneous speech in
a highly noisy environment for the RATS corpus \cite{sadjadi2013unsupervised} and then optimized for the Apollo data in \cite{ziaei2014speech}.
In short, five-dimensional feature vectors including four different speech voicing measures, obtained in time or frequency domains, plus a perceptual spectral flux feature
are transformed by Principle Component Analysis (PCA) to one-dimensional Combo features. 
A two-Gaussian GMM is then fit to the Combo features supposing each Gaussian represents one of the speech or non-speech classes. 
As the amount of speech and non-speech data in the Apollo corpus is not balanced, an alternative model of speech using data from 
a separate corpus is build
and embedded within the Combo-SAD framework in \cite{ziaei2014speech} which is considered as the baseline in this challenge. 

\begin{figure}[t!]
  \centering
  \includegraphics[width=0.48\textwidth]{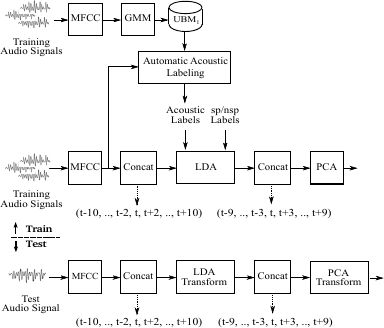}
  \caption{Block diagram of the temporal context-based feature transformation in the EML Online SAD algorithm.}
  \label{fig:EML-SAD-Front-End}
\end{figure} 

\section{EML Online SAD Algorithm}
\label{sec:proposed-algorithm}
In summary, typical spectral features are first extracted and transformed into a more speech/non-speech discriminative domain. 
Then two types of SAD vectors are extracted every 10 frames (0.1 sec) given a Universal Background Model (UBM). 
The SAD vectors are compared with adaptive SAD vectors from the model and the decisions are made based on an adaptive threshold
at runtime. More details are given as follows. 

\subsection{Feature Extraction}
\label{subsec:feature-extraction}
Feature vectors are 12 dimensional Mel-Frequency Cepstral Coefficients (MFCCs) extracted with a 25~msec window every 
10~msec. Filter banks are started from 150 Hz. MFCCs are mean normalized over a 1 sec sliding window and then appended with $\Delta$ and 
$\Delta\Delta$ coefficients.

\subsection{Temporal Context-Based Feature Transformation}
\label{subsec:Context-Based-Feature-Transformation}
In order to increase the discriminative power of feature vectors, they are transformed in two stages given the temporal
contexts as in Fig.~\ref{fig:EML-SAD-Front-End}. The first stage is supervised and based on a Linear Discriminative Analysis (LDA)
which is trained given automatically created acoustic labels and the provided speech/non-speech (sp/nsp) labels.
Acoustic labels are created given a Gaussian Mixture Model (GMM) which is trained using unlabeled training data and is supposed
to model the entire acoustic space. This UBM is referred to as UBM$_{1}$ in Fig.~\ref{fig:EML-SAD-Front-End} and in the experiments. The idea is that 
each Gaussian in the GMM represents an acoustic class and each acoustic class could be shared between speech and non-speech frames
mainly due to highly noisy data.
Therefore, LDA could help to discriminate speech and non-speech frames in the acoustic class level given the temporal context.
As an example, if a GMM with 64 Gaussian is fit to the training data, LDA will be trained with at most 128 classes
(64 acoustic $\times$ 2 sp/nsp).
Note that some acoustic classes could represent only speech or non-speech frames.
The temporal context considered for this stage is $(t-10,..,t-2,t,t+2,..,t+10)$, meaning that these frames are concatenated before
LDA training.

The second stage is unsupervised and based on a PCA to take advantage of all other unlabeled data.
PCA will be applied on the outputs of the first stage and with a different context configuration as
$(t-9,t-6,t-3, t,t+3,t+6,t+9)$.

\subsection{Speech Activity Detection Vectors}
\label{subsec:SAD-vectors}
Two types of low dimensional vectors are used in this work to represent speech and non-speech segments. 
The first one is based on zero-order Baum-Welch statistics extracted from a GMM-based UBM
as in~\cite{ghahabi_ESSV_VAD_2018}. In summary, given the UBM, a single vector of zero-order statistics is extracted from 
training speech frames and another from non-speech frames, representing the modeling SAD vectors.
One difference with \cite{ghahabi_ESSV_VAD_2018} is that for UBM training and statistics 
extraction, the transformed features from Sec.~\ref{subsec:Context-Based-Feature-Transformation} are used rather than MFCCs. 
Another difference is that two GMMs with the same size are fit separately to speech and non-speech data and then they 
are merged to build the final UBM, referred to as UBM$_{2}$ in the experimental section. The experimental results showed that both changes significantly boost the speech detection 
accuracy. More details about Baum-Welch statistics and UBM can be found in \cite{reynolds2009gaussian,ghahabi_ESSV_VAD_2018}.

The second type of vectors are based on the centralized first-order Baum-Welch statistics which will build 
supervectors. Supervectors are then transformed by a DNN which is trained to discriminate speech and non-speech 
short duration segments. The final SAD embedding vectors are extracted from one of the hidden layers.
The first-order Baum-Welch statistics are extracted given a similar UBM and feature vectors as in the first type of vectors mentioned above.
The only difference is that in this case
the UBM is smaller in order to keep the dimensions of supervectors 
as low as possible and because statistics are extracted from only 10 or 20 frames. 
This UBM will be referred to as UBM$_{3}$ in section \ref{sec:results}.
The speech and non-speech classes will then be represented by two single embedding vectors obtained by averaging 
over training speech and non-speech embeddings. 
These embeddings are used as complementary vectors to make more robust and reliable decisions. 


In the testing phase, the two types of SAD vectors are first extracted for an unknown short duration segment
and the resemblance ratio score will be based on 
the cosine similarity of the test vector 
with each modeling SAD vector as follows,
\begin{equation} 
S_{sp}(\boldsymbol w_{unk}^{\mathtt T,i})=\cos(\boldsymbol w_{unk}^{\mathtt T,i},\boldsymbol w_{sp}^{\mathtt M})-\cos(\boldsymbol w_{unk}^{\mathtt T,i},\boldsymbol w_{nsp}^{\mathtt M})
\end{equation}
where $\boldsymbol w_{unk}^{\mathtt T,i}$ is the $i$th unknown test vector, $S_{sp}(.)$ is the score of being a speech and not a non-speech segment, $\boldsymbol w_{sp}^{\mathtt M}$ 
and $\boldsymbol w_{nsp}^{\mathtt M}$ are, respectively, speech and non-speech vectors from the model.

\subsection{Model and Threshold Adaptations}
\label{subsec:adaptation}
We consider two buffer windows for the test vectors detected in the past as speech and non-speech with 
the lengths of $L_{sp}$ and $L_{nsp}$, respectively. Then the SAD vectors from the model and the decision threshold are updated after each 
decision as follows,
\begin{equation} 
\boldsymbol w_{sp}^{\mathtt A,i}=(1-\alpha)\boldsymbol w_{sp}^{\mathtt M} + \frac{\alpha}{L_{sp}}\sum_{j=1}^{L_{sp}} \boldsymbol w_{sp}^{\mathtt T,j} 
\label{eq:sp-adaptation}
\end{equation}
\begin{equation} 
\boldsymbol w_{nsp}^{\mathtt A,i}=(1-\alpha)\boldsymbol w_{nsp}^{\mathtt M} + \frac{\alpha}{L_{nsp}}\sum_{k=1}^{L_{nsp}} \boldsymbol w_{nsp}^{\mathtt T,k} 
\label{eq:nsp-adaptation}
\end{equation}
\begin{equation} 
\theta^{\mathtt A,i}=(1-\beta)\theta^{\mathtt M} + \frac{\beta}{L_{sp}}\sum_{j=1}^{L_{sp}} S_{sp}(\boldsymbol w_{sp}^{\mathtt T,j} )
\label{eq:th-adaptation}
\end{equation}
where $\boldsymbol w_{sp}^{\mathtt A,i}$ and $\boldsymbol w_{nsp}^{\mathtt A,i}$ are
the adapted speech and non-speech models for the $i$th testing vector, $\theta^{\mathtt A,i}$ is the adapted 
threshold in that point, $\theta^{\mathtt M}$ is the tuned threshold for the non-adapted models. 
$\boldsymbol w_{sp}^{\mathtt T,j}$ and $\boldsymbol w_{nsp}^{\mathtt T,k}$ are 
the test vectors in the buffers which have been detected as speech and non-speech segments in the past, and
$\alpha$ and $\beta$ are the adaptation weights for the models and the threshold, respectively.
In other words, the adapted SAD models will be the weighted sum of the fixed models and the average over the last detected 
speech or non-speech vectors. 
The threshold is also adapted based on a weighted sum of the fixed threshold $(\theta^{\mathtt M})$ and the average over the last 
$L_{sp}$ scores from the segments detected as speech.

\section{Experimental Results}
\label{sec:results}
We have used only the sp/nsp labeled data provided in the challenge. Only the train set is used for training of all 
supervised and unsupervised parts of the algorithm. The dev set is used only for monitoring the DNN training and for tuning the parameters.
The feature dimensions after LDA and PCA transformations (Sec.\ref{subsec:Context-Based-Feature-Transformation}) 
are 12 and 24, respectively. The size of UBM$_{1}$ for automatic acoustic labeling is 32. For the first type of 
SAD vectors based on zero-order statistics, two GMMs with 64 Gaussian each, 
resulting in a UBM with 128 Gaussian (UBM$_{2}$), are fit on speech and non-speech frames. The size of UBM$_{3}$ for extracting the
first-order Baum-Welch statistics
is set to 32. 

The DNN used in these experiments is a Multilayer Perceptron (MLP) with low-dimensional ($24\times32$) 
supervectors as inputs and two nodes as output layer representing the sp/nsp classes. 
The cross-entropy is used as the objective function and Variable ReLU~\cite{ghahabi2018restricted} is used 
as the activation function. We tried different configurations and analyzed the embedding qualities extracted from different layers.
The experimental results showed that the top layers overfit so quickly and the only reliable layer was the first layer 
right above the input layer. So, we trained a DNN with 3 hidden layers and the first hidden layer 
was used for embedding extraction.
Moreover, we analyzed the embeddings after each epoch and picked epoch 30 for the final 
evaluation as it was providing the most complementary information to the first type of SAD vectors.

The buffer window lengths for speech and non-speech segments 
($L_{sp}$ and $L_{nsp}$ in Eqs.~\ref{eq:sp-adaptation}-\ref{eq:th-adaptation}) are set to 30 
and 60 segments corresponding to 3 and 6 sec, respectively. The adaptation weights $\alpha$ and $\beta$ 
(Eqs.~\ref{eq:sp-adaptation}-\ref{eq:th-adaptation})
are set to 0.4 and 0.1, respectively. The decision threshold for the non-adaptive system ($\theta^{\mathtt M}$) is set to 0.25. 
As the scores for both types of SAD vectors are based on $\cos$ similarity and they are normalized against the non-target
class (non-speech in this case), they are naturally normalized and will be always in a range between $-2$ and $2$. 
So, we just use the average of scores obtained from the two types of SAD vectors at runtime.

Table~\ref{table:DCF-results} compares the best configuration of the EML online SAD system with the baseline results provided by the challenge 
organizers and the prior work of the authors \cite{ghahabi_ESSV_VAD_2018} in which MFCCs without any transformation are used 
as feature vectors and no model or threshold adaptation technique is applied. 
As it can be seen, the results are much better with a big margin compared to both baseline results.
As the hypothesized system outputs on the evaluation set are evaluated 
only by the challenge organizers and due to the limited number of 
daily submissions, the results for the second baseline system on the eval set is not available in Table~\ref{table:DCF-results}.

\begin{table}[t!]
\centering
\begin{threeparttable}[b]
  \footnotesize
  \setlength{\tabcolsep}{12.8pt}
 \renewcommand{\arraystretch}{1.1}
 \caption{DCF (in \%) comparison of the EML online SAD system with the challenge baselines on the Fearless Step Challenge Phase III.
 A 0.25 sec collar is considered for evaluations.}
 \label{table:DCF-results}
\begin{tabular}{lcc}
\hline
\textbf{\begin{tabular}[c]{@{}l@{}}\end{tabular}} & \textbf{Dev} & \textbf{Eval} \\ \cline{1-3} 
Challenge Baseline \cite{joglekar2020fearless}                                                       & 12.50           &  	15.61          \\ \hline
EML Online Baseline \cite{ghahabi_ESSV_VAD_2018}                                                       & 5.14           &  	-          \\ \hline
EML Online Algorithm                                                         & \textbf{2.20}              &   \textbf{3.80}            \\ \hline 
\end{tabular}
\end{threeparttable}
\end{table}

Experimental results showed that the adaptation proposed in Eqs.~\ref{eq:sp-adaptation}-\ref{eq:th-adaptation}
improves the accuracy by about 10\% and 5\% on 
the dev and eval sets, respectively. We did not observe any noticeable improvement by adapting the embedding-based models. This could be due to 
some overfitting but it needs more investigation. Therefore, we have adapted
only the SAD vectors based on zero-order statistics for the final results. We also noticed that the dev set is too small
to reliably tune the parameters or conclude something specific. So, at most of the time, we used both the train and dev sets 
for this purpose even though it is not so realistic. 
Moreover, our analysis showed that the embedding-based vectors did not perform better than the zero-order-based vectors 
in general but they helped to make more robust decisions specifically for utterances with very low SNRs. 

For feature transformation, we also tried to use only PCA, only LDA, or first PCA and then LDA. All these configurations
performed worse than the one proposed in Sec.\ref{subsec:Context-Based-Feature-Transformation}, i.e., first LDA and then PCA.
The reason could be LDA at the first point can reduce the dimension of concatenated feature vectors more efficiently and 
PCA-transformed features will be more compatible with GMM training at the end. 

One of the main advantages of the proposed system is that it works online with a very low computational cost.
The real-time factor of the whole speech activity detection process is about 0.002 using a single Intel Xeon CPU with 2.30 GHz.
This makes it possible to process a large collection of data with reasonable accuracy and speed. 

\section{Conclusions and Future Work}
\label{sec:conclusions}
  
This paper presented the EML online algorithm for the
phase III of the Fearless Steps challenge. The proposed algorithm is a combination
of both supervised and unsupervised techniques and
assigns speech and non-speech labels at runtime every approximately
0.1 second. Compared to our previous algorithm, feature
vectors are transformed in two stages given the temporal contexts
and a model and threshold adaptation method is used for
increasing the robustness of the algorithm. The experimental results
showed a competitive accuracy on both development and
evaluation datasets with a real-time factor of about 0.002 using
a single CPU machine.  Future work may address the incorporation of the
unlabeled data which has not been used in the experiments reported in this paper.

\section{Acknowledgements}
Part of the work  presented in this paper has been funded by the German Federal Ministry of Education and Research within the AdaptAR project (Grant no. 02K18D073).

\balance
\bibliographystyle{MyIEEEtran}

\bibliography{Mybib}


\end{document}